\documentclass[a4paper]{article}
\usepackage{CJK}   
\usepackage{indentfirst}  
\usepackage{bm}    
\usepackage{graphicx}  
\usepackage{epsfig} 
\usepackage{amsmath}  

\begin{document}    

\begin{CJK*}{GBK}{song}  



\begin{center}
\LARGE\bf Scaling behaviour of magnetic entropy change in bilayered manganites
by two-variable polynomials fitting to magnetization $^{*}$   
\end{center}

\footnotetext{
\hspace*{-.45cm}\footnotesize $^*$
Project supported by National
Natural Science Foundation of China (Grant No. 61565013) and The Scientific Research Project of the Inner Mongolia Autonomous Region Colleges and Universities (Grant No. NJZZ199).}
\footnotetext{
\hspace*{-.45cm}\footnotesize $^\dag$ 
Corresponding author. E-mail: xubao79@163.com }

\begin{center}
\rm Bao Xu $^{\rm a)\dagger}$ 
\end{center}

\begin{center}
\begin{footnotesize} \sl
${}^{\rm a)}$ Key Laboratory of Magnetism and Magnetic Materials at Universities of Inner Mongolia Autonomous Region and Department of Physics Science and Technology,
Baotou Teachers' College, Baotou, 014030 \\   
\end{footnotesize}
\end{center}


\vspace*{2mm}

\begin{center}
\begin{minipage}{15.5cm}
\parindent 20pt\footnotesize
Based on the two-variable polynomial model of magnetization,
magnetic entropy change of bilayered manganites with $327$-structure
and its scaling behaviour with respect to applied magnetic fields are investigated.
It's found that the Curie temperature, which is defined as the point
at which the partial derivative of magnetization with respect to temperature
reaches its maximum, is different from the temperature of peak magnetic entropy change.
Thus a mean-field model can not apply to this kind of manganites.
In contrast to what has been found in manganites with the $113$-structure,
the scaling behaviour at the Curie temperature in manganites with $327$-structure
is much different from that at the temperature of peak magnetic entropy.
It's also found that the temperature dependence of the scaling exponent under weak fields
is distinct from that under strong fields.
This difference is attributed to an crossover from one-step transition under weak fields
to two-step transition under strong fields.
\end{minipage}
\end{center}

\begin{center}
\begin{minipage}{15.5cm}
\begin{minipage}[t]{2.3cm}{\bf Keywords:}\end{minipage}
\begin{minipage}[t]{13.1cm}
scaling exponent, magnetic entropy, bilayered manganite
\end{minipage}\par\vglue8pt
{\bf PACS: }
75.40.Cx, 75.30.Sg, 75.60.Ej
\end{minipage}
\end{center}

\section{Introduction}  

For magnetic materials,
both physical properties coupled with the magnetic degree of freedom
and associated microscopic coupling mechanism
can be revealed from magnetization data.
Following are some examples:
First of all, the magnetic transition points,
and the type of effective exchange couplings under molecular field approximation (ferromagnetic, antiferromagnetic, or ferrimagnetic type),
can be determined from the temperature dependence of inverse susceptibility.
The second example is to identify the order of magnetic transitions according to
the Banerjee criterion, subsequently qualitatively assess
the strength of spin-lattice couplings,
and eventually provide valuable information for
the construction and test of different microscopic models.
The third one mentioned here is to estimate the grain distribution of
polycrystalline, especially nanocrystalline, samples
through investigating the ratio of spontaneous magnetization at finite temperature with respect to
that at zero point.
By comparing the resultant distribution with that obtained from X-ray diffraction (XRD),
one can give clues to optimizing the chemical doping
and also controlling the preparation process.
The fourth one is for first-order transition materials.
By measuring its first-order reversal curve (FORC),
the contribution to magnetization from irreversible rotations
can be quantitatively analysed, and therefore information about magnetocrystal anisotropy and
associated lattice symmetry is provided.
The fifth but not the last one is to calculate magnetic entropy change
from magnetization data
and assess its potential value as a new magnetocaloric material.

In spite so much information is revealed by magnetization data,
the analysis results quite depend on the particular model and approximation
that are employed.
This paper aims at studying the the scaling properties of
magnetic entropy change $\Delta S_{H}$ based on series approximation of magnetization.
The magnetic entropy change at temperature $T$ and maximum applied field $H$ is estimated as
\begin{equation}
\Delta S_{H} = \int_{0}^{H}\left(\frac{\partial M}{\partial T}\right)_{H^{\prime}} dH^{\prime}. \label{entr1}
\end{equation}
Its scaling exponent with respect to applied magnetic field $n$ is defined as
\begin{eqnarray}
n &=& \frac{ \partial \ln |\Delta S_{H}| }{ \partial \ln H } \label{n0} \\
&=& \frac{H}{\Delta S_{H}}\left(\frac{ \partial  M }{ \partial  T }\right)_{H}. \label{n1}
\end{eqnarray}
In deriving expressions (\ref{n0}) and (\ref{n1}),
the magnetic entropy at $H=0$ with fixed $T$ is assumed to be a constant number.
and the Maxwell relation $\left(\frac{ \partial  S_{H} }{ \partial  H }\right)_{T}
=  \left(\frac{ \partial  M }{ \partial  T }\right)_{H}$
is used.

To calculate exponent $n$,
the model most often used is the approximated version of molecular field theory at $M$
much smaller than its saturated value. Under this approximation,
the equation of state can be expressed as$^{[1,2]}$
\begin{equation}
\frac{H}{M} = a(T-T_{\rm C}) + b M^{2} \label{eos0}
\end{equation}
and exponent $n$ can be easily obtained as$^{[3]}$ 

\begin{equation}
n(T) = \frac{  2M^{2} }{  3M^{2}-M_{s}^{2}  }\theta(T\leq T_{\rm C})
+  \frac{2H/M}{H/M + 2bM^{2}}\theta(T > T_{\rm C})  \label{n3}
\end{equation}
where, spontaneous magnetization $M_{s}$ equals
$\left[a(T_{\rm C}-T)/b\right]^{1/2}$ at temperatures lower than $T_{\rm C}$;
$\theta(A)=1,0$ correspond to logical expression $A$ being true and false,
respectively. Note that $n=1,\frac{2}{3},2$ in cases with
$T\ll T_{\rm C}$ ($M$ approximately equal to $M_{s}$),
$T\to T_{\rm C}$ ($M_{s}$ approaching zero) and
$T\gg T_{\rm C}$ ($M^{2}$ being a small quantity), respectively$^{[4]}$.
The above approximation applies to
materials with weak short-range correlations.

When short-range correlations becomes important,
the molecular field model (\ref{eos0}) fails to give correct
results in the vicinity of critical point $T_{\rm C}$.
In this case one can employ the semi-empirical
Arrott-Noakes equation of state$^{[5]}$
\begin{equation}
\left(\frac{H}{M}\right)^{1/\gamma} = a(T-T_{\rm C}) + bM^{1/\beta}, \label{eos1}
\end{equation}
with $a$, $b$, $T_{\rm C}$, $\beta$ and $\gamma$ are parameters to be determined.
Near to the critical point, $\beta$ and $\gamma$ correspond to
critical exponents defined as
$M_{s} \propto (T_{\rm C}-T)^{\beta}$ and
$ \left( \frac{\partial M}{\partial H} \right)_{H\to 0} \propto (T-T_{\rm C})^{\gamma} $, respectively.
It's considered that the Arrott-Noakes equation is
particularly suitable to deal with magnetization data in the vicinity of the critical point.
From equation (\ref{eos1}), it's easy to obtain following relations at $T_{\rm C}$


\begin{equation}
\left( \frac{\partial M}{\partial T} \right)_{H}
= -  \frac{a}{b} \frac{\beta\gamma}{\beta+\gamma} M^{ 1-\frac{1}{\beta} }
\propto H^{ \frac{\beta-1}{\beta+\gamma} }, \label{akeqn1}
\end{equation}
and
\begin{equation}
\Delta S_{H} = - \frac{a}{b^{1-\gamma}} \frac{\beta\gamma}{2\beta+\gamma-1}
M^{2+\frac{\gamma-1}{\beta}}
\propto H^{\frac{2\beta+\gamma-1}{\beta+\gamma}}. \label{akeqn2}
\end{equation}

Substituting (\ref{akeqn2}) into (\ref{n0}), exponent $n$ can be obtained as$^{[6,7]}$
\begin{equation}
n (T_{\rm C}) = 1 + \frac{\beta - 1}{\beta+\gamma}  \label{n4}.
\end{equation}
The amazing thing is that,
by substituting (\ref{akeqn1}), (\ref{akeqn2}) into (\ref{n1}),
exponent in (\ref{n4}) is again obtained.
By assigning $\gamma =1$ and $\beta=1/2$, equation (\ref{n4})
reduces to the molecular field result $2/3$.
It's also noted that the scaling exponents of $\Delta S_{H}$ with respect to magnetic fields
equals that of $H\left(\frac{\partial M}{\partial T}\right)_{H}$.

To determine $n$ at $T_{\rm C}$, one can perform fitting to
the field dependence of peak magnetic entropy $\Delta S_{H}^{\rm pk}(H)$.
This method does not strictly distinguish $T_{\rm C}$
from the temperature of peak magnetic entropy, $T_{\rm pk}$.
For estimating exponent $n$ within the whole measuring temperatures,
it's usually to utilize the definition in (\ref{n1}).
This paper not only uses above two methods but also
apply (\ref{akeqn1}) to estimating $n$ from magnetization data.

The paper is organized as follows:
Section 2 provides the model and formula used in this work;
In Section 3 the numerical results and associated discussions can be found.
Conclusions are put into Section 4.

\section{Model and method}

\subsection{Series estimation of magnetization without symmetry restriction}

Before doing the series estimation, make size transformations as follow:
Applied magnetic field $H$, temperature $T$ and magnetization $M$ are given as

\begin{eqnarray*}
H &=& f_{H}(x) = H_{\rm min} + x(H_{\rm max} - H_{\rm min}), \\
T &=& f_{T}(y) = T_{\rm min} + y(T_{\rm max} - T_{\rm min}), \\
M &=& f_{M}(z) = M_{\rm min} + z(M_{\rm max} - M_{\rm min}),
\end{eqnarray*}
where, dimensionless variants $x,y,z \in [0,1]$ represent
reduced magnetic field, temperature and magnetization.
For brevity, the $\mu_{0}$ before $H$ is omitted and
the applied fields are measured in Tesla.


Giving up the symmetry restriction connecting magnetization and applied fields,
the reduced magnetization $z$ can be represented as a series in $x$ and $y$
\begin{equation}
z = \sum_{t=0}^{S}c_{t}h_{t}(x,y) \label{se1}
\end{equation}
where $h_{t}(x,y) = 1,x,y,x^{2},xy,y^{2},x^{3},\cdots$ and
$S$ denotes the maximum index in above series expression.
To determine coefficients $c_{t}$,
a straight method is two-variable orthogonal polynomial fitting to experimental data.

\subsection{Estimation of coefficients $c_{t}$}

Expression (\ref{se1}) can be estimated by rewriting it in
normalized orthogonal polynomials $P_{s}(x,y)$ as
\begin{equation}
\hat{z}(x,y) = \sum_{t=0}^{S} b_{t}P_{t}(x,y),\label{fit_expr}
\end{equation}
where, $P_{t}(x,y)$ is the $t$-th orthogonal polynomial
and $b_{t}$ the corresponding coefficient with subscripts $t\geq 0$.

By applying the Gram-Schmidt orthogonalization process
to linearly independent functions $h_{t}(x,y)$,
the orthogonal polynomial $P_{s}(x,y)$ can be recursively generated as
\begin{equation*}
P_{s}(x,y)= a_{ss} h_{s}(x,y) + \sum_{t=0}^{s-1}a_{st} P_{t}(x,y).
\end{equation*}
Coefficients $a_{ss}$ and $a_{st}$ are given by summing over
$N$ experimentally recorded values, $(x_{i},y_{i},z_{i})$, as
\begin{equation}
a_{ss} = - \left[ \sum_{i=1}^{N} P_{0}(x_{i},y_{i})h_{s}(x_{i},y_{i}) \right]^{-1}, \label{e2}
\end{equation}
\begin{equation}
a_{st} = - a_{ss} \sum_{i=1}^{N} P_{t}(x_{i},y_{i}) h_{s}(x_{i},y_{i}). \label{e3}
\end{equation}

By minimizing the fitting error with a regularization term,
characterized by parameter $\lambda$, as
\begin{equation}
\sigma_{1} = \frac{1}{N} \sum_{i=1}^{N}\left[  \hat{z}(x_{i},y_{i})-z_{i} \right]^2
+ \lambda \left[ \nabla^{2} \hat{z}(x_{i},y_{i}) \right]^{2}, \label{fit_eqn}
\end{equation}
coefficient $b_{t}$ is determined as
\begin{equation}
b_{t} =
\frac{
- \lambda R_{t} Q_{t} + \sum_{i=1}^{N}z_{i} P_{t}(x_{i},y_{i})
}{
1 + \lambda  \left( Q_{t} \right)^{2}  },			\label{fit_coef_1}
\end{equation}
where,

\begin{eqnarray*}
R_{t} &=& \sum_{r=0}^{t-1}b_{r}Q_{r},\\
Q_{t} &=& \sum_{i=1}^{N}\nabla^{2} P_{t}(x_{i},y_{i}),
\end{eqnarray*}
with $t\geq 0$.
After $a_{st}$ and $b_{s}$ being determined,
the values of coefficient $c_{t}$
can be readily computed from them.

\subsection{Spontaneous magnetization,
partial derivative of magnetization with respect to temperature,
and magnetic entropy change }

By expressing subscript $t$ as
\begin{equation}
t(m,j)=\frac{1}{2}m(m+1)+j  \label{indx}
\end{equation}
with $m\geq 0$ and $0\leq j\leq m$,
reduced magnetization $z$ in (\ref{se1}) can be expressed as
\begin{equation}
z = \sum_{p=0}^{p_{S}} z_{p} x^{p} \label{fit_expr}
\end{equation}
where
\begin{equation}
z_{p} = \sum_{j=0}^{j_{p}} c_{t(j+p,j)} y^{j},
\end{equation}
and $p_{S}$ and $j_{p}$ are maximum indices of $p$ and $j$, respectively.

\subsubsection{Spontaneous magnetization}

Spontaneous magnetization at reduced temperature $y$ is expresses as
\begin{equation}
M_{0} = f_{M} (z_{0})
\end{equation}
with
\begin{equation}
z_{0} = \sum_{j=0}^{j_{0}} c_{t(j,j)} y^{j}.
\end{equation}
Other coefficients before the powers
like $x^{j}$ with $1\leq j \leq j_{p}$ can be similarly obtain.

\subsubsection{Partial derivative of magnetization with respect to temperature}

At fixed magnetic field $H$, the partial derivative of $M$ with respect to $T$ can be estimated as

\begin{equation}
\frac{ \partial  M }{ \partial  T }
= A_{1} \frac{ \partial  z(x,y) }{ \partial  y }
= A_{1} \sum_{t=0}^{S} c_{t}  \frac{ \partial  h_{t}(x,y) }{ \partial  y }, \label{mdt}
\end{equation}
with
\begin{equation}
A_{1} = ( M_{\rm max} - M_{\rm min} ) / (T_{\rm max} - T_{\rm min} ).
\end{equation}

Representing $t$ as in (\ref{indx}), the summation over $t$ in (\ref{mdt})
can be rewritten as

\begin{equation}
\sum_{t=0}^{S} c_{t(m,j)}   \frac{ \partial  h_{t(m,j)}(x,y) }{ \partial  y }
=\frac{1}{x}\sum_{k=1}^{k_{\rm S}} \epsilon_{k}(y) x^{k}  \label{mdt1}
\end{equation}
where,
\begin{equation}
\epsilon_{k}(y) = \sum_{j=1}^{j_{k}} c_{t(k+j-1,j)}\cdot j\cdot y^{j-1}, \label{epsilon}
\end{equation}
\begin{equation}
j_{k} = \left\lfloor \sqrt{2S+2k+\frac{1}{4}} -k -\frac{1}{2} \right\rfloor,
\end{equation}
and
\begin{equation}
k_{S} =\left\lfloor \sqrt{2S-\frac{7}{4}} - \frac{1}{2} \right\rfloor,
\end{equation}
with $\lfloor \cdot \rfloor$ denoting the rounding down operation.
In above derivations, index replacement $m = k+j-1$ has been used.


\subsubsection{Magnetic entropy change}

The magnetic entropy change is calculated according to expression (\ref{entr1})

\begin{equation}
\Delta S_{H} =
A_{2}\int_{0}^{x} \frac{\partial z(x^{\prime},y)}{\partial y} {\rm d} x^{\prime}.
\end{equation}
In above equation, the integral can be computed as
\begin{equation*}
\int_{0}^{x}\frac{\partial z(x^{\prime},y)}{\partial y} {\rm d} x^{\prime}
= \sum_{t=0}^{S} c_{t}
\int_{0}^{x}\frac{\partial h_{t}(x^{\prime},y)}{\partial y}{\rm d}x^{\prime}.
\end{equation*}
Thus the magnetic entropy change can be estimated as
\begin{equation}
\Delta S_{\rm M} = A_{2} \sum_{t=0}^{S} c_{t}
\int_{0}^{x}\frac{\partial h_{t}(x^{\prime},y)}{\partial y}{\rm d}x^{\prime}, \label{mtrp}
\end{equation}
where the leading factor $A_{2}$ is expressed as
\begin{equation*}
A_{2} = (M_{\rm max} - M_{\rm min})(H_{\rm max} - H_{\rm min})/(T_{\rm max} - T_{\rm min}).
\end{equation*}
At fixed reduced temperature $y$, equations (\ref{mtrp}) can be rewritten as

\begin{equation}
\sum_{t=0}^{S} c_{t(m,j)} \int_{0}^{x}
\frac{ \partial  h_{t(m,j)}(x^{\prime},y) }{ \partial  y } {\rm d}x^{\prime}
=\sum_{k=1}^{k_{\rm S}}\frac{1}{k} \epsilon_{k}(y) x^{k} \label{mtrp1}
\end{equation}

By comparing equation (\ref{mdt1}) and (\ref{mtrp1}),
it's easy to see that $T_{\rm C}$
determined from the maximum magnitude
of partial derivative $\big| \left(\frac{\partial M}{\partial T}\right)_{H} \big|_{\rm max} $,
generally differs from $T_{\rm pk}$ that is identified from the peak value of
magnetic entropy change $| \Delta S_{H}^{\rm pk} |$; and
the difference increases with the applied field.
Actually, inspections of the recursive relations (\ref{zdy}) and (\ref{rec_entropy})
give the following relation

\begin{equation}
\int_{0}^{x} \frac{ \partial  h_{t(m,j)}(x^{\prime},y) }{ \partial  y } {\rm d}x^{\prime}
= \frac{x}{m-j+1}\frac{ \partial  h_{t(m,j)}(x,y) }{ \partial  y }.
\end{equation}

\subsection{Scaling exponent of magnetic entropy change with respect to applied magnetic fields}

We next calculate exponent $n$.
Substituting (\ref{mdt}) and (\ref{mtrp}) into (\ref{n1}),
exponent $n$ can be expressed as
\begin{equation}
n = \frac{A_{1}}{A_{2}} \frac{f_{H}(x)}{x}
\frac{   \sum_{t=0}^{S} c_{t}  x\frac{ \partial  h_{t}(x,y) }{ \partial  y }
}{
\sum_{t=0}^{S} c_{t} \int_{0}^{x}
\frac{ \partial  h_{t}(x^{\prime},y) }{ \partial  y } {\rm d}x^{\prime}   }   \label{n5}
\end{equation}

By fixing $H_{\rm min}=0$, the leading factor in (\ref{n5}) equal to $1$,
we now reach the expression of exponent $n$ as
\begin{equation}
n = \frac{ \epsilon_{1}(y)  + \epsilon_{2}(y) x + \epsilon_{3}(y) x^{2} + \cdots
+ \epsilon_{k_{S}}(y) x^{k_{S}-1}    }{
\epsilon_{1}(y)  + \frac{1}{2}\epsilon_{2}(y) x + \frac{1}{3}\epsilon_{3}(y) x^{2}
+ \cdots + \frac{1}{k_{S}} \epsilon_{k_{S}}(y) x^{k_{S}-1} } \label{n6}
\end{equation}

Note that $n$ depends on the reduced temperature at fixed reduced field.
For the exponent $n$ in (\ref{n3}),
it's obvious that at weak magnetic fields, i.e.,
much smaller than saturation field and $x$ being small quantity,
the exponent $n\approx 1$.
At $T$ much higher than $T_{\rm C}$, the $k=1$ term should be abandoned for
the vanishing spontaneous magnetization,
and exponent $n$ approaches $2$ at weak magnetic fields.
It's noted that in the range of $T\ll T_{\rm C}$ or $T\gg T_{\rm C}$,
exponent $n$ is not dependent on $y$, i.e., reduced temperature.
When the applied field increases up to the saturation field, i.e., $x\to 1$,
exponent $n$ is expressed as
\begin{equation}
n = \frac{ \epsilon_{1}(y)  + \epsilon_{2}(y)  + \epsilon_{3}(y)  + \cdots
+ \epsilon_{k_{S}}(y)     }{
\epsilon_{1}(y)  + \frac{1}{2}\epsilon_{2}(y)  + \frac{1}{3}\epsilon_{3}(y)
+ \cdots + \frac{1}{k_{S}} \epsilon_{k_{S}}(y) }.
\end{equation}

\subsection{Exponent $n$ at $T_{\rm C}$ and $T_{\rm pk}$}

When exponent $n$ at the transition point is considered,
the situation becomes complicated as $T_{\rm C}$ generally differers from $T_{\rm pk}$.
$T_{\rm C}$ is determined in this work as the temperature
where the maximum value of the partial derivative of magnetization with respect to temperature,
$\big|\left(\frac{\partial M}{\partial T}\right)_{H}\big|_{\rm max}$, occurs.
Exponent $n$ at $T_{\rm C}$ can be determined as
\begin{equation}
\Big|\left(\frac{\partial M}{\partial T}\right)_{H}\Big|_{\rm max} \propto H^{n-1}. \label{ntc}
\end{equation}
$T_{\rm pk}$ is the temperature at which the peak value
of magnetic entropy change, $\Delta S_{H}^{\rm pk}$, appears.
Exponent $n$ as $T_{\rm pk}$ is determined as
\begin{equation}
\big|\Delta S_{H}^{\rm pk}\big| \propto H^{n}. \label{ntpk}
\end{equation}
It needs stressing that both (\ref{ntc}) and (\ref{ntpk})
are derived from the Arrott-Noakes equation (\ref{eos1}), with
assumption that parameters $a$, $b$, $T_{\rm C}$, $\beta$ and $\gamma$
do not depend on temperatures or applied fields.
Another point needs noticing is that exponent $n$
at $T_{\rm C}$ and $T_{\rm pk}$ are usually considered to the same.

\subsection{Useful recursive relations}

\subsubsection{$h(i)\leftarrow\frac{\partial h_{i}(x,y)}{\partial y}$
used to determine partial derivative of magnetization with respect to temperature}

\begin{eqnarray}
&&h(0) = 0;~~ h(1) = 0;~~ h(2) = 1;~~s=1;
\nonumber\\
&&\mathrm{For}~~ m\geq 2
\nonumber\\
&& \{ s = s + m;
\nonumber\\
&&h(s+j) = \left\{
\begin{array}{cl}
0, & j=0;\\
x\cdot h(s+j-m), & 1\leq j \leq m-1;\\
\frac{m}{m-1}\cdot y\cdot h(s-1) , & j=m;
\end{array}
\right.
\nonumber\\
&&m = m + 1.\}  \label{zdy}
\end{eqnarray}

\subsubsection{$h(i)\leftarrow\int_{0}^{x}\frac{\partial h_{i}(x^{\prime},y)}{\partial y}{\rm d}x^{\prime}$
used to compute magnetic entropy change}

\begin{eqnarray}
&&h(0) = 0;~~ h(1) = 0;~~ h(2) = x;~~s=1;
\nonumber\\
&&\mathrm{For}~~ m\geq 2
\nonumber\\
&& \{ s = s + m;
\nonumber\\
&&h(s+j) = \left\{
\begin{array}{cl}
0, & j=0;\\
\frac{m-j}{m-j+1}\cdot x \cdot h(s+j-m), & 1\leq j \leq m-1;\\
\frac{m}{m-1}\cdot y \cdot h(s-1), & j=m;
\end{array}
\right.
\nonumber\\
&&m = m + 1.\}    \label{rec_entropy}
\end{eqnarray}

\section{Results and discussions}


Here, we apply above method to deal with the magnetization data of polycrystalline samples
La$_{1.2}$Sr$_{1.8}$Mn$_{2}$O$_{7}$ obtained with Physical Property Measurement System
(PPMS) of Quantum Design Company. More details can be found in reference $^{[8]}$. 
Following calculations use the whole data ($3789$) as the training group.
The regularization parameter $\lambda=e^{-24}\approx 0.3775{\times}10^{-10}$,
is selected out by comprehensive considerations about the
overfitting degree $\gamma$ (with sampling factor equal to $3$) and
magnetic entropy change $|\Delta S_{H}|$.
With $96$ orthogonal polynomials ($S=95$),
the fitting error reached is $0.34540{\times}10^{-4}$.
For comparison, the fitting error with $S=0$ is  $0.78397{\times}10^{-1}$.

Fig. 1 summarizes the temperature dependence of magnetic entropy change
under applied magnetic fields $0.25 {\rm T}\leq H\leq 5.00 {\rm T}$.
Calculations are according to the recursive relations defined in (\ref{mtrp}).
To find out the present value of $\lambda$,
we have also created magnetization data on the uniform mesh
from the fitted expression (\ref{fit_expr});
and compared the value of $|\Delta S_{H}|$, estimated according to equation (\ref{mtrp}),
with that calculated according to finite difference
\begin{equation}
\Delta S_{H}\left(H_{j_{\rm max} },\frac{T_{i}+T_{i+1}}{2}\right)
= - \sum_{j=1}^{j_{\rm max}-1} \frac{ M(H_{j+1},T_{i+1}) - M(H_{j+1},T_{i}) }{ T_{i+1} - T_{i} }(H_{j+1}-H_{j}).
\end{equation}
At the value of $\lambda$ given above,
no obvious fluctuations appear in the vicinity of $T_{\rm pk}$,
which manifests that overfitting is not significant.

Fig. 2 displays the temperature dependence of spontaneous magnetization.
Note that in contrast to the $113$-structure,
spontaneous magnetization in bilayered manganites
does not approaches zero at the Curie temperature $T_{\rm C}^{0}$ at vanishing magnetic fields,
which is considered to be the temperature at which spontaneous magnetization
reaches the minimum.
The fact that the minimum is not equal to zero at $T_{\rm C}^{0}$,
might be attributed to the finite magnetization in the Mn-O two-layers above $T_{\rm C}^{0}$.
Therefore, a two-step magnetization process takes place in the bilayered manganites.
Since experimental recorded data are fewer in this temperature range,
the fitting error at $T \geq 135 {\rm K}$ is obviously larger than the global error,
and therefore we cannot identify the temperature, at which the spontaneous magnetization equals zero,
to be the real phase transition point.

Fig. 3 compares the magnetic-field dependence of Curie temperature ($T_{\rm C}$)
with that of the temperature of peak magnetic entropy change ($T_{\rm pk}$).
It's noted that $T_{\rm C}$ is bigger than $T_{\rm pk}$ in bilayered manganites.
This result is in contract to that found in reference,$^{[7]}$
the latter announces that $T_{\rm pk}$ is equal to $T_{\rm C}$
while using the mean field model,
and larger than $T_{\rm C}$ with the Heisenberg model.
It's also noted that with increasing field intensity,
the difference $T_{\rm C}-T_{\rm pk}$ increases up to the maximum
and then lowers down to negative value at about $\mu_{0}H = 5$ T.
The strange field-dependence of $T_{\rm C}$ might suggest
spin dimerization occurs within the Mn-O two-layers,
as similar field-dependences are usually
found in spin-dimerized insulators.
However, it needs to stress that the electronic itinerating properties and
magnetic frustrations in the bilayered manganite might mask
the step-by-step magnetization that is found in other spin-dimerized two-layered compounds.
Hence, the perfect step structure of magnetization
can not be observed in two-layered manganite.

Shown in Fig. 4 are the magnetic field dependences of magnetic entropy change $\Delta S_{H}$,
the partial derivatives of magnetization with respect to temperature
$\Big|\left(\frac{\partial M}{\partial T}\right)_{H}\Big|$, and
$|H\left(\frac{\partial M}{\partial T}\right)_{H}|$,
at $T_{\rm C}$ and $T_{\rm pk}$.
The magnetic entropy $\Delta S_{H}$ is measured in ${\rm J}\cdot {\rm Kg}^{-1}\cdot {\rm K}^{-1}$
and magnetization $M$ in ${\rm A}\cdot {\rm m}^{2}\cdot {\rm kg}^{-1}$.
We try to determine the scaling exponents $n$ at $T_{\rm C}$ and $T_{\rm pk}$
according to
$H|\left(\frac{\partial M}{\partial T}\right)_{H}|_{\rm max}\propto H^{n}$ and
$|\Delta S_{H}^{\rm pk}|\propto H^{n}$.
It's noted that two distinct dependences of $|H\left(\frac{\partial M}{\partial T}\right)_{H}|$
on the applied field at $H\leq 1.75 {\rm T}$ and $H\geq 1.75 {\rm T}$.
We find that $n(T_{\rm C}) = 1$ with $H\leq 1.75 {\rm T}$
and $n(T_{\rm C})\approx 0$ with $H \geq 1.75 {\rm T}$ can explain the obtained results.
Hence, the crossover between two different scaling laws happens at bout $H = 1.75 {\rm T}$.
For the magnetic entropy change, however, we can not
find a proper numerical value of $n(T_{\rm pk})$
to unify the estimated results in the whole range of measurement.
In contrast, a nonlinear dependence is found like
\begin{equation}
\ln|\Delta S_{H}^{\rm pk}| = 0.19 + 0.99 \ln H  - 0.16 \ln^{2} H.
\end{equation}
It needs emphasizing that under weak fields, exponent $n$ equal to $1$ is
a direct consequence of the non-vanishing spontaneous magnetization at
$T_{\rm C}$ or $T_{\rm pk}$,
which is the typical characteristic of bilayered compounds.
Comparing panel (a) with (c), it is noted that the value of exponent $n$ estimated according to
(\ref{n1}) is unreasonably large at $T_{\rm C}$, and that $T_{\rm pk}$ seems more reliable.

\begin{figure}
\includegraphics[ scale=0.3 ]{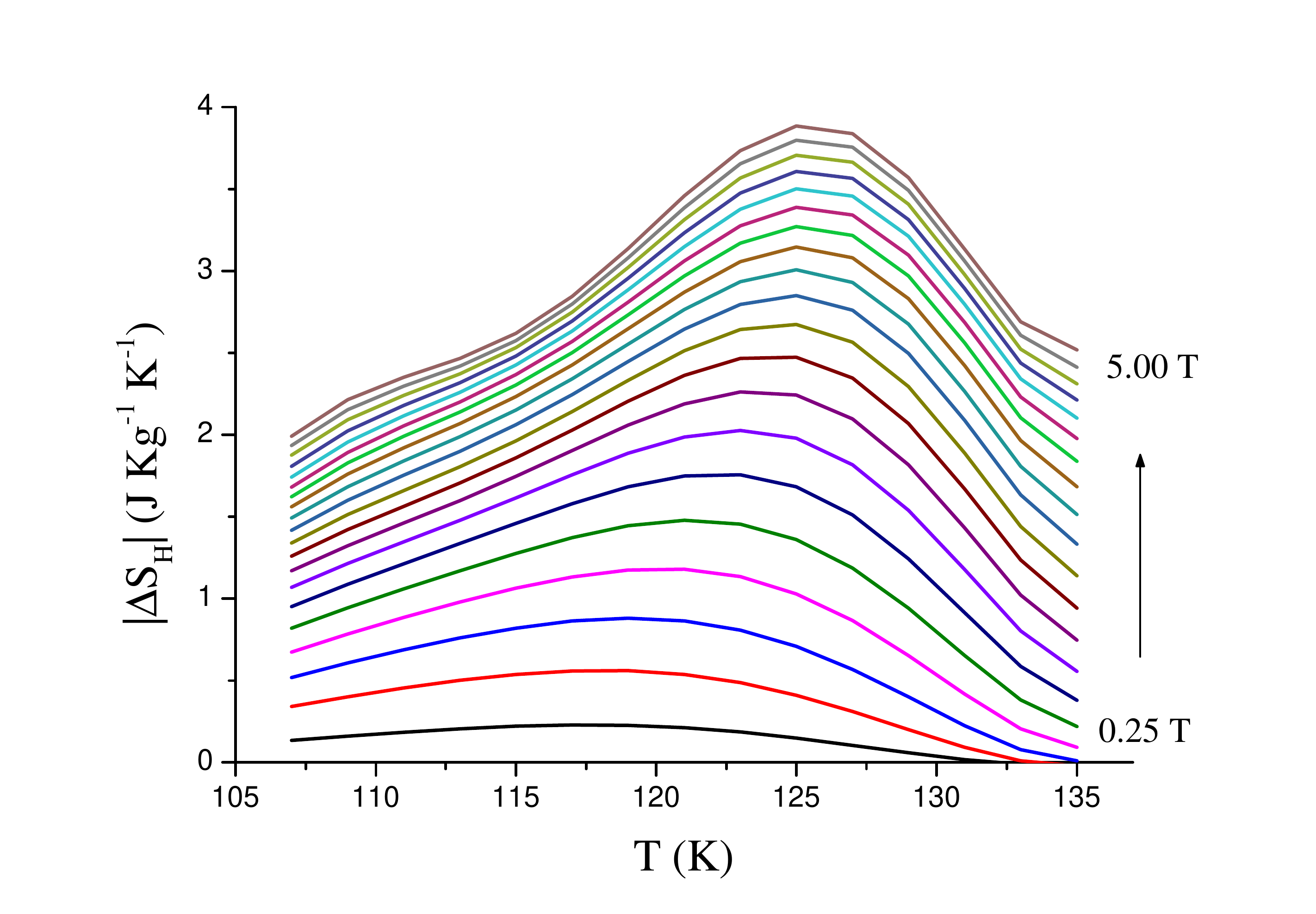}
\caption{
The temperature dependence of magnetic entropy change under
applied magnetic fields $0.25 {\rm T}\leq H\leq 5.00 {\rm T}$.
}\label{fig1}
\end{figure}

\begin{figure}
\includegraphics[ scale=0.3 ]{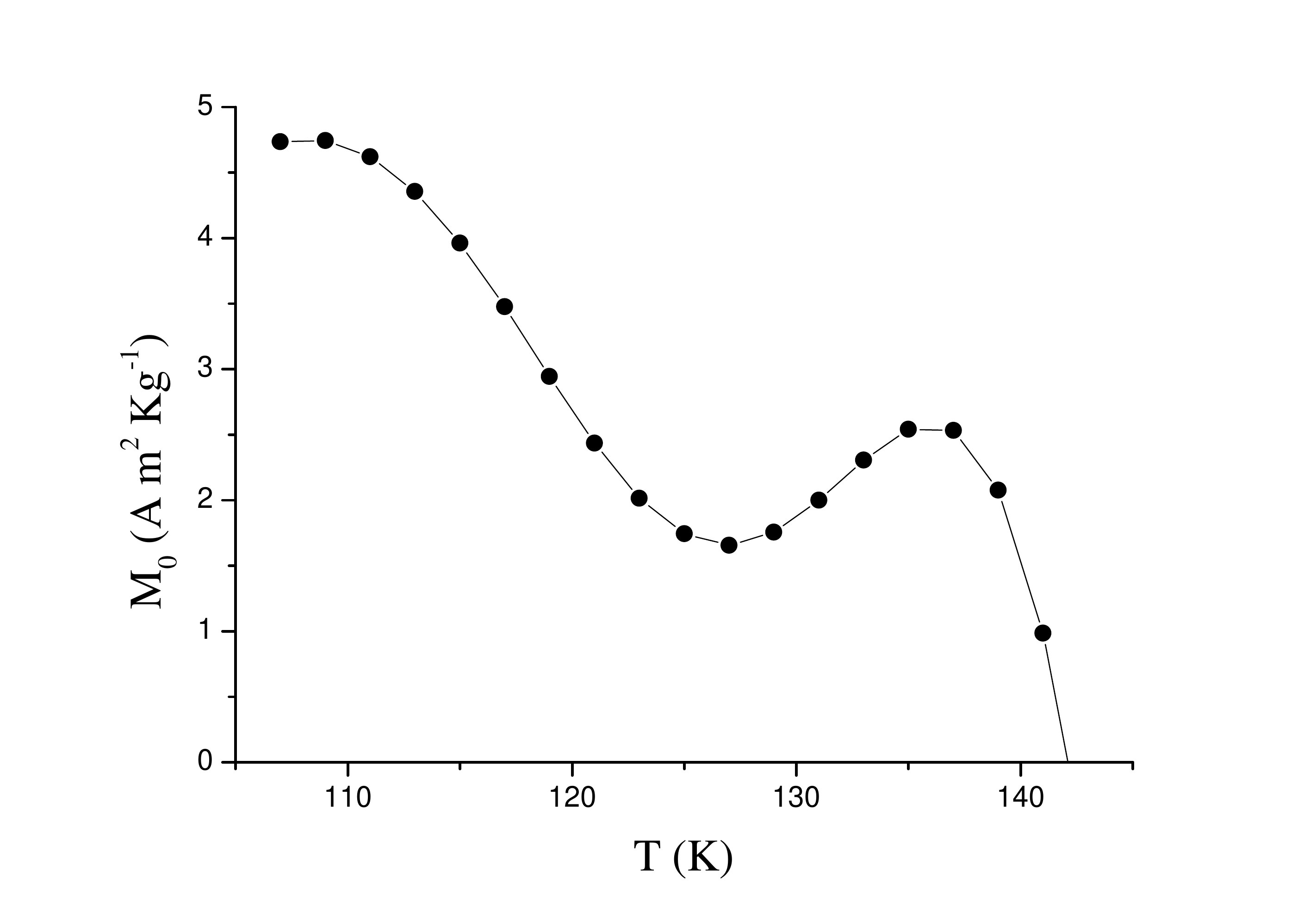}
\caption{
The temperature dependence of spontaneous magnetization.
}\label{fig2}
\end{figure}

\begin{figure}
\includegraphics[ scale=0.3 ]{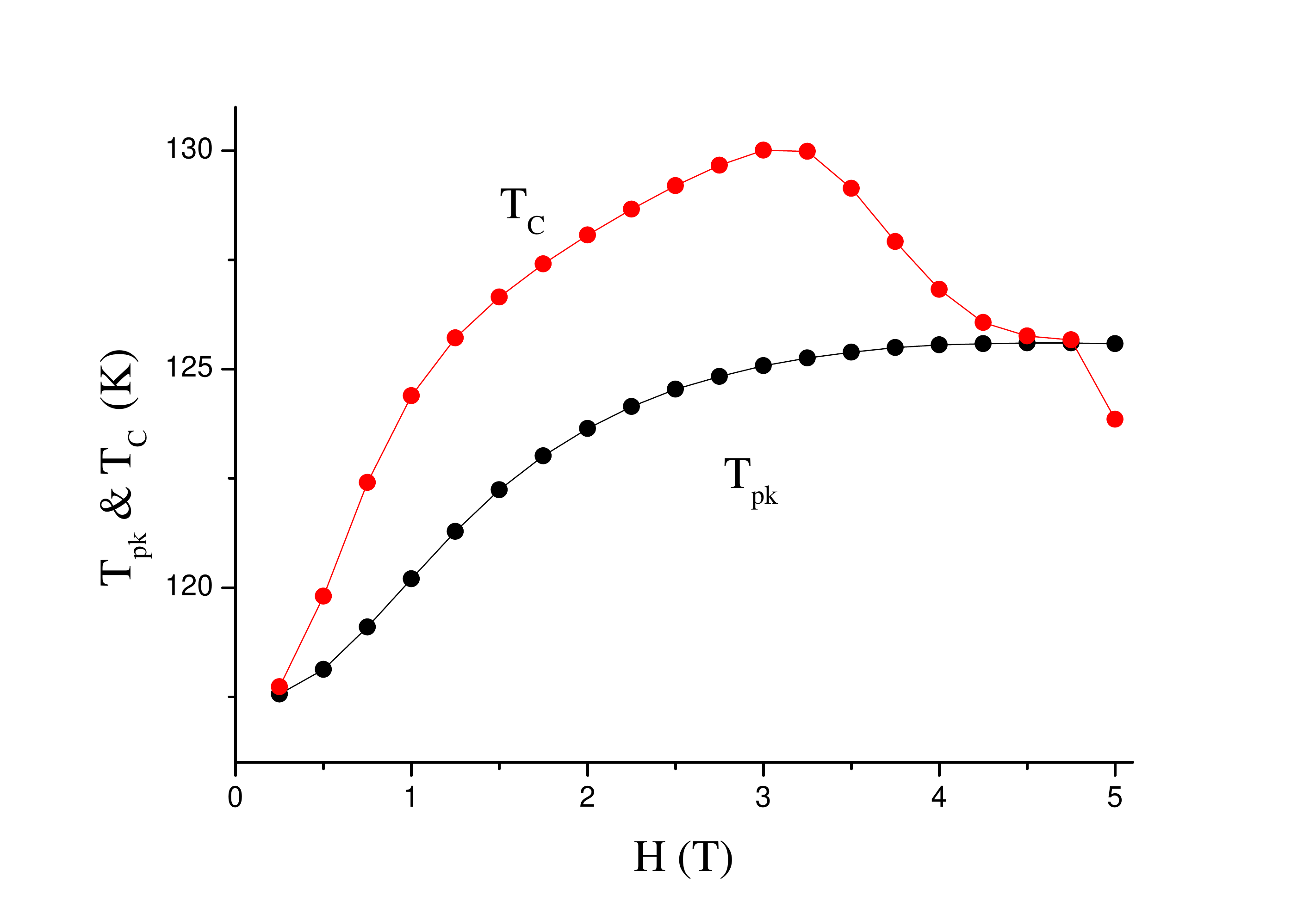}
\caption{
Comparison between the magnetic-field dependence of Curie temperature ($T_{\rm C}$)
and that of the peak temperature of magnetic entropy change ($T_{\rm pk}$).
$T_{\rm C}$ is determined as the temperature at which
$\Big|\left(\frac{\partial M}{\partial T}\right)_{H}\Big|$ reaches its maximum.
The strange field-dependence of $T_{\rm C}$ might suggest
spin dimerization occurs within the Mn-O two-layers.
}\label{fig3}
\end{figure}

\begin{figure}
\includegraphics[ scale=0.5 ]{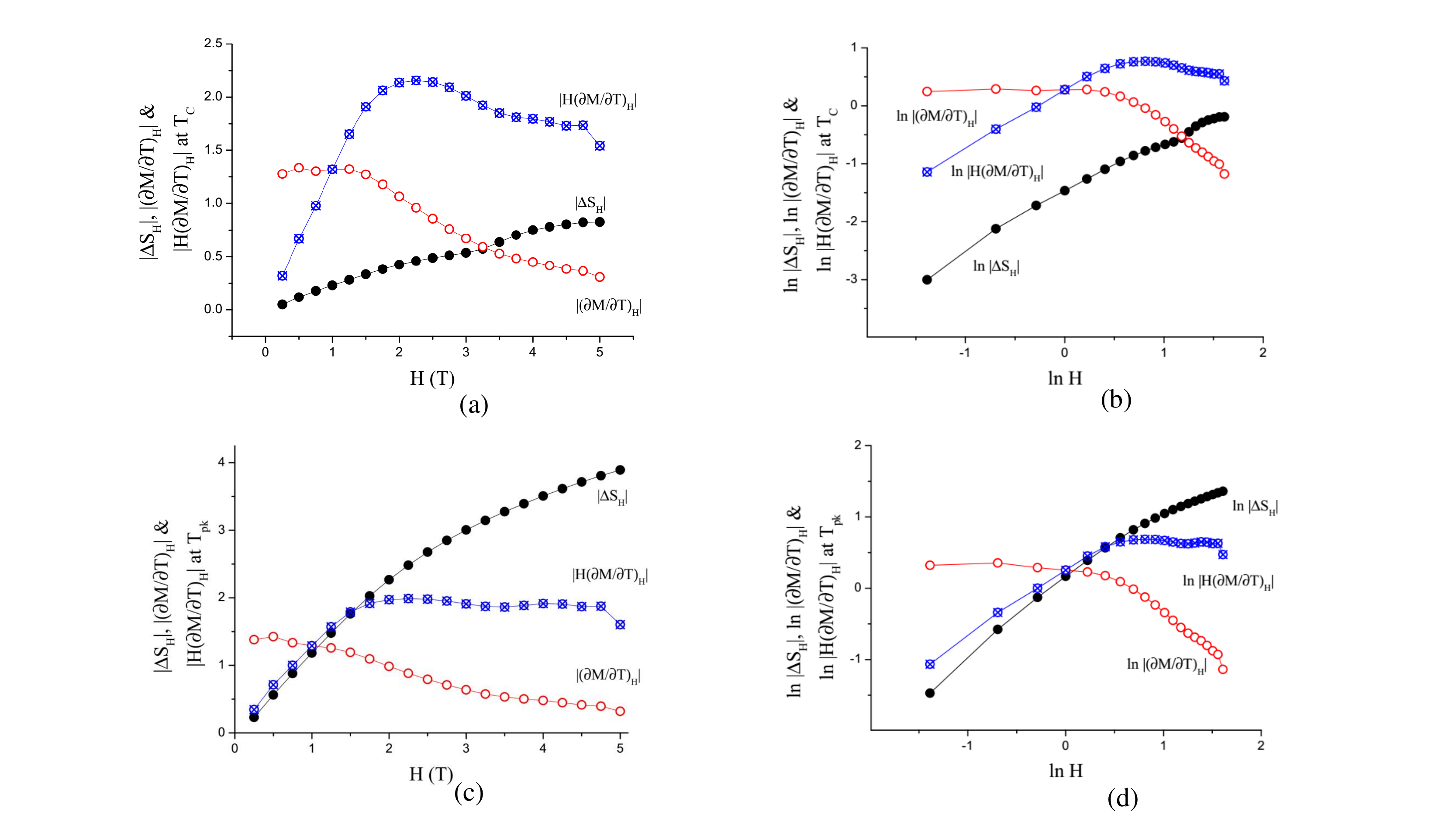}
\caption{
The magnetic field dependence of magnetic entropy change $\Delta S_{H}$,
the partial derivatives of magnetization with respect to temperature
$\Big|\left(\frac{\partial M}{\partial T}\right)_{H}\Big|$, and
$|H\left(\frac{\partial M}{\partial T}\right)_{H}|$,
at $T_{\rm C}$ and $T_{\rm pk}$.
$T_{\rm C}$ is defined as the temperature at which
$\Big|\left(\frac{\partial M}{\partial T}\right)_{H}\Big|$ reaches its maximum,
and $T_{\rm pk}$ the temperature at which
$|\Delta S_{H}|$ reaches its peak value.
The magnetic field is measured in ${\rm Tesla}$,
$\Delta S_{H}$ in ${\rm J}\cdot {\rm Kg}^{-1}\cdot {\rm K}^{-1}$
and magnetization $M$ in ${\rm A}\cdot {\rm m}^{2}\cdot {\rm kg}^{-1}$.
}\label{fig4}
\end{figure}

Shown in Fig. 5 is
the temperature dependence of
the scaling exponent of magnetic entropy change with respect to applied magnetic field.
It is noticed that two different transitions seem present under strong magnetic fields.
It may be explained by one transition in the vicinity of the transition point occurs
under weak magnetic fields, and a magnetic structure rearrangement and
therefore two-step process happens under strong fields.

\begin{figure}
\includegraphics[ scale=0.3 ]{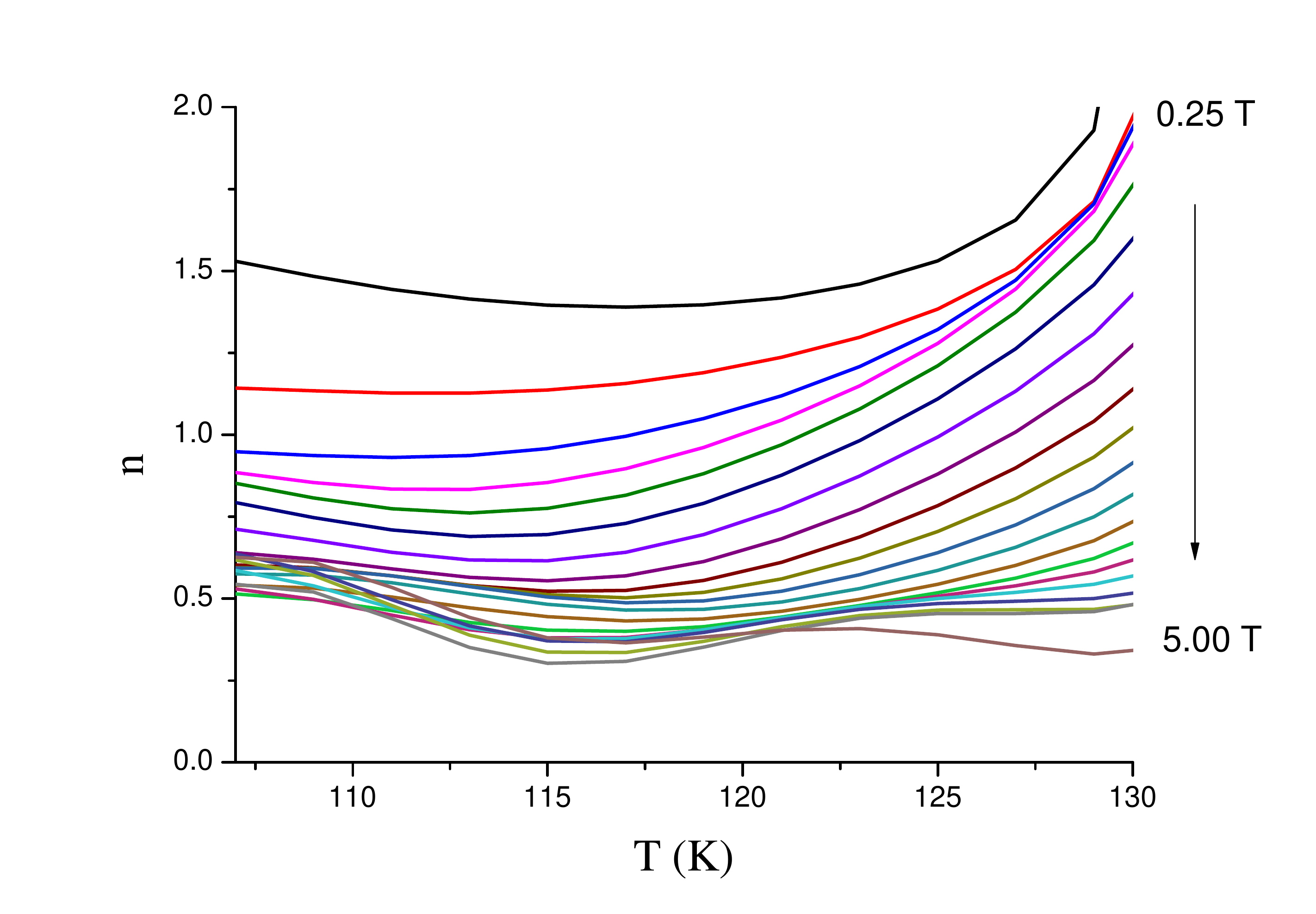}
\caption{
The temperature dependence of
the scaling exponent of magnetic entropy change with respect to applied magnetic field.
}\label{fig5}
\end{figure}

\section{Conclusions}

In conclusion, we apply an two-variable polynomial fit to the magnetization
of bilayered manganites, in order to investigate its magnetic entropy change
and the associated scaling behaviour with respect to applied magnetic fields.
It's found that the Curie temperature is different from the temperature of peak magnetic entropy change.
The difference between these two temperatures are dependent on applied magnetic fields.
Therefore a mean-field theory does not apply to bilayered manganites.
The field dependence of the Curie temperature might imply weak dimerization occurs
in the bilayered manganite.
In contrast to what has been found in manganites with the 113 structure,
the scaling behaviour at the Curie temperature in bilayered manganites
is much different from that at the peak temperature.
Hence it requires distinguishing the actual transition point from the peak temperature while discussing the scaling law of magnetic entropy change.
It's also found that the temperature dependence of the scaling exponent at peak temperatures
under weak fields is distinct from that under strong fields.
This difference is attributed to an crossover from one-step transition under weak fields
to two-step transition under strong fields.
To further test the validity of the method provided in this paper,
other kinds of magnetic materials will be analysed in the future.


\end{CJK*}  
\end{document}